# Throughput-Distortion Computation Of Generic Matrix Multiplication: Toward A Computation Channel For Digital Signal Processing Systems

Davide Anastasia and Yiannis Andreopoulos[*]

**ABSTRACT**

The generic matrix multiply (GEMM) function is the core element of high-performance linear algebra libraries used in many computationally-demanding digital signal processing (DSP) systems. We propose an acceleration technique for GEMM based on *dynamically* adjusting the *imprecision* (distortion) of computation. Our technique employs adaptive scalar companding and rounding to input matrix blocks followed by two forms of packing in floating-point that allow for concurrent calculation of multiple results. Since the adaptive companding process controls the increase of concurrency (via packing), the increase in processing throughput (and the corresponding increase in distortion) depends on the input data statistics. To demonstrate this, we derive the optimal throughput-distortion control framework for GEMM for the broad class of zero-mean, independent identically distributed, input sources. Our approach converts matrix multiplication in programmable processors into a *computation channel*: when increasing the processing throughput, the output noise (error) increases due to *(i)* coarser quantization and *(ii)* computational errors caused by exceeding the machine-precision limitations. We show that, under certain distortion in the GEMM computation, the proposed framework can significantly surpass 100% of the peak performance of a given processor. The practical benefits of our proposal are shown in a face recognition system and a multi-layer perceptron system trained for metadata learning from a large music feature database.

## 1. INTRODUCTION

*Can we accelerate basic mathematical operations in computers by systematically trading off precision in the results in a stochastic manner?* We address this question for one of the most essential operations in digital signal processing (DSP) systems: matrix multiplication.

[*] The authors are with the University College London, Dept. of Electronic & Electrical Engineering, Torrington Place, WC1E 7JE, London, UK; Tel: +44-20-76797303 (both authors); fax: +44-20-73889325 (both authors); email: d.anastasia@ee.ucl.ac.uk (D. Anastasia), iandreop@ee.ucl.ac.uk (Y. Andreopoulos). The work was supported by EPSRC, project EP/F020015/1.
[*]Corresponding author. Address and contact information: see above.

Computationally-intensive DSP operations, such as covariance scatter matrix calculation [1], noise cancellation [2], back-propagation learning [3], and two-dimensional (2D) transform analysis and synthesis of large data sets [4] make heavy use of matrix multiplication operations. Optimized realizations of such operations in programmable processors are based on the standard basic linear algebra subprograms (BLAS) library[1], which is tailored to the particular hardware via the use of optimized assembly or single-instruction-multiple-data (SIMD) code [5]. Single and double-precision floating-point matrix multiplication is realized in BLAS by the generic matrix multiply (GEMM) function. Since most BLAS functions can be rewritten to use GEMM as the dominant operation as the problem size scales [6], GEMM throughput measurements have traditionally been considered important enough to form a core part of processor benchmarking efforts.

A. *Related Work*

Most practical approaches for complexity-distortion scalability are *algorithm-specific* [7]-[10]. Although significant complexity reduction can be offered by such proposals, it is difficult to provide dynamic precision adaptation; essentially, one is forced to adopt an algorithmic modification leading to certain "generally acceptable" precision, without being able to adjust this at run-time, under the same implementation, and in an inexpensive manner.

Within more generic, theory-driven, approaches, Monte Carlo algorithms have been proposed for fast approximate matrix multiplication [11]. The concept of approximate or stochastic computation was proposed as a broader way to achieve complexity-distortion scaling [8] and fault tolerance [12]. Such approaches target error-tolerant applications where the aim is to provision for *average performance vs. distortion* requirements and *not for the worst case*. However, all existing proposals [8][11][12] are either based on complexity models or customized VLSI designs, and cannot be deployed in programmable processors that have fixed hardware for arithmetic, logic and memory units. One exception is work on software designs for image processing operations, where progressive (incremental) computation for general-purpose or DSP systems was achieved [13][14]. However, the proposed schemes apply only for lossy-to-lossless image processing operations and, importantly, no distortion-controlled acceleration of linear operations is proposed.

---

[1] E.g.: www.netlib.org (Netlib repository), www.intel.com/software/products/mkl (Intel Math Kernel Library), http://developer.amd.com/cpu/Libraries/acml (AMD Core Math Library), http://www.maplesoft.com/ (Maple), http://www.mathworks.com/ (Matlab), http://eigen.tuxfamily.org/ (Eigen – ported to digital signal processors like the ARM Neon, the TI TMS320C6x series, etc.), provide optimized realizations of BLAS.

*B. Proposed Work*

In this paper we focus on high-performance realizations of the GEMM function of BLAS [5][6] (reviewed in Section 2). Our proposal for GEMM centers on: *(i)* SIMD-based realization of scalar quantization of inputs; *(ii)* a novel form of input packing for concurrent calculation of multiple results beyond what is possible with SIMD instructions (Section 3). Based on the detailed analysis of the quantization-induced error and the representation-induced error (Section 4), a throughput/distortion control mechanism is proposed for each subblock of GEMM (Section 5 and 6). Depending on the distortion tolerance or the throughput requirement for GEMM, the experiments of Section 7 demonstrate that our design ranges from zero distortion, i.e. identical throughput/precision to the conventional single or double-precision GEMM (`sGEMM` or `dGEMM`, respectively), up to 180% higher throughput with controlled distortion in the results. Since the throughput/distortion adaptation depends on the input statistics and computational errors can be introduced by exceeding the machine precision limitations, this system can be seen as a computation channel for high-performance matrix multiplication in programmable processors. Finally, even though the proposed control framework is tailored to the broad class of zero-mean, independent identically distributed (iid) input data, with some loss of optimality we also validate it in applications were inputs are not strictly iid (Section 8). Such applications are: a state-of-the-art face recognition system based on the 2D principal component analysis (PCA) algorithm [1] and multilayer perceptron-based training [3] applied to automated metadata identification from music features [15].

This work extends our initial efforts [16] by:

*(i)* proposing different ways for packing of quantized input data and comparing their performance;

*(ii)* incorporating the impact of the representation-induced noise in the final distortion;

*(iii)* extending the initial control framework for the adaptation of GEMM toward the general case of multiple packings and providing a solution for *dual* control based on throughput or distortion;

*(iv)* presenting extended experimental results that include both single and double-precision representations and extending these results within several DSP applications.

As such, this paper and the accompanying demonstration materials [17] present the first generic approach toward systematic throughput/precision realizations of matrix-based DSP operations in programmable processors. This links this proposal with error-tolerant computing [18] and the imminent scaling and power-wall problems of programmable processors [19], with the added benefit of being directly applicable in today's computing infrastructure without requiring architectural or hardware modifications.

## 2. OVERVIEW OF GEMM

Consider the GEMM design depicted in Figure 1(a), following the general flow found in optimized BLAS implementations [5]. The application calls `sGEMM` or `dGEMM` for an $M \times K$ by $K \times N$ matrix multiplication[2], [top-level of Figure 1(b)] which is further subdivided into $L \times L$ "inner-kernel" matrix multiplications. For our approach, $L$ is specified by:

$$L = \frac{\text{SIMD}_{\text{bytes}}}{b_{\text{repr}}} \times k \times \text{LCM}(2,3,\ldots,\max\{W\}) \tag{1}$$

with: $\text{LCM}(a_1, \ldots, a_N)$ the least common multiple of integers $a_1, \ldots, a_N$; $\max\{W\}$ the maximum number of packings utilized in the proposed throughput/distortion computation; $\text{SIMD}_{\text{bytes}}$ the number of bytes of each SIMD register ($\text{SIMD}_{\text{bytes}} = 16$ in this work); $b_{\text{repr}} \in \{4,8\}$ the number of bytes for single and double-precision floating point representations; and $k$ any positive integer. The *inner-kernel* result $\mathbf{R}_{2,1}$ of the example shown in Figure 1(a) comprises multiple *subblock* multiplications $\mathbf{A}_{2,l}\mathbf{B}_{l,1}$:

$$\mathbf{R}_{2,1} = \sum_{l=0}^{K/L-1} \mathbf{A}_{2,l}\mathbf{B}_{l,1}. \tag{2}$$

If the matrices' dimensions are not multiples of $L$, some "cleanup" code [5] is applied at the borders to complete an inner-kernel result of the overall matrix multiplication.

This separation into top-level processing and subblock-level processing is done for efficient cache utilization. Specifically, during the initial data access of GEMM for top-level processing, data in matrix **A** and **B** is reordered into block major format: for each $L \times L$ pair of subblocks $\mathbf{A}_{i,l}$ and $\mathbf{B}_{l,j}$ multiplied to produce inner-kernel result $\mathbf{R}_{i,j}$, $0 \leq l < \frac{K}{L}$, $0 \leq i < \frac{M}{L}$, $0 \leq j < \frac{N}{L}$, the input data within $\mathbf{A}_{i,l}$ and $\mathbf{B}_{l,j}$ is reordered in rowwise and columnwise raster manner, respectively. Thus, sequential data accesses are performed during the actual computation and this enables the use of SIMD instructions for each subblock multiplication, thereby leading to significant acceleration.

As shown in Figure 1(b), our proposal creates an intermediate level (Tier 1.5) that performs certain pre-processing for companding, rounding and packing before calling the subblock matrix multiplication code. Once the calculation is completed for each subblock multiplication, post-processing is applied to retrieve the results. The subblock-specific adjustment of the companding factors in our approach allows for data-driven (adaptive) packing and concurrency in the calculation of the inner-kernel results – thus, `sGEMM` and `dGEMM` are accelerated according to the input statistics of each subblock. Since our modifications are applied externally to the $L \times L$ subblock matrix multiplication, in the next section we focus on the subblock processing stage. If the matrix sizes are not multiples of $L$, trivial size modifications are required for the cleanup code at the borders [5], which are omitted for brevity of description.

---

[2] Notations: Boldface uppercase and lowercase letters indicate matrices and vectors, respectively; the corresponding italicized lowercase indicate their individual elements, e.g. **A** and $a_{m,n}$; all indices are integers; max{**A**} and min{**A**} indicate the maximum and minimum values of matrix **A** and $\|\mathbf{A}\|_F$ is the Frobenius (root-sum-square) norm; $[\![x]\!]$ rounds $x$ to integer and $\lfloor x \rfloor$ is the largest integer not greater than $x$; $\tilde{x}, \bar{x}, \hat{x}$ indicate the quantized, packed, and dequantized value of $x$, respectively; all random variables used in this paper have mean zero and are represented by Greek lowercase letters, e.g. $\alpha \sim \text{P}_\alpha(\sigma_\alpha)$, with $\text{P}_\alpha$ their zero-mean probability density function with standard deviation $\sigma_\alpha$ (P and $\sigma$ are reserved for this purpose).

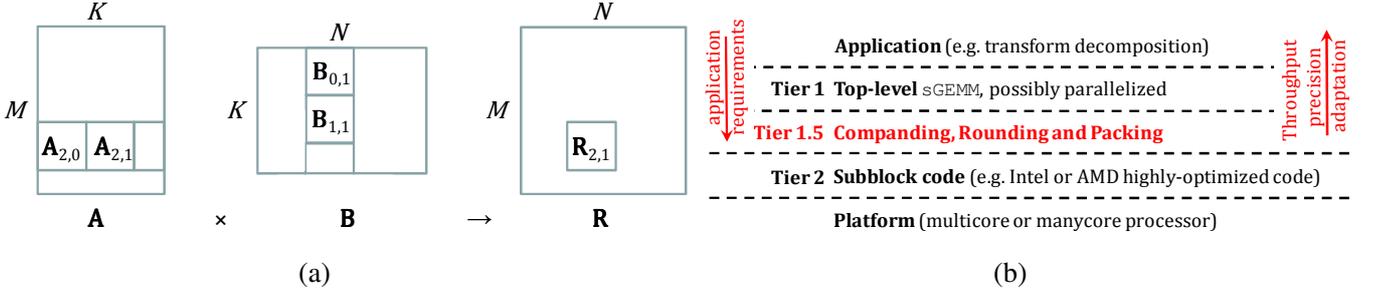

**Figure 1. (a)** Illustration of outer and inner-kernel processing of GEMM in BLAS; **(b)** positioning of the proposed approach within the execution environment of signal processing tasks based on optimized linear algebra libraries.

## 3. PROPOSED APPROACH

Our proposal is adjusting the quantization of the inputs according to their numerical range. As such, for each $L \times L$ inner-kernel $\mathbf{R}_{i,j}$ we need to calculate the maximum and minimum values within each subblock $\mathbf{A}_{i,l}$ and $\mathbf{B}_{l,j}$ [Figure 1(a), $0 \leq l < \frac{K}{L}$], and store them for the companding stage. These calculations are performed during the data reordering stage and thus do not require additional data accesses. Since the remainder of this section is focusing on subblock-level operations, for notational simplicity, we remove indices from subblocks $\mathbf{A}_{i,l}$ and $\mathbf{B}_{l,j}$ and, despite the block-major format reordering, we retain the 2D indexing in the equations.

### A. *Scalar Quantization via Companding and Rounding*

We perform uniform companding and rounding within each subblock $\mathbf{A}$ and $\mathbf{B}$, assuming input ranges $[\min\{\mathbf{A}\}, \max\{\mathbf{A}\}]$ and $[\min\{\mathbf{B}\}, \max\{\mathbf{B}\}]$. For any element $a_{m,n}, b_{m,n}$ of $\mathbf{A}, \mathbf{B}$, $0 \leq m, n < L$, we have:

$$\tilde{a}_{m,n} = [\![c_{\mathbf{A}} a_{m,n}]\!], \quad \tilde{b}_{m,n} = [\![c_{\mathbf{B}} b_{m,n}]\!]. \tag{3}$$

Hence, $\tilde{a}_{m,n}$ are integers within $\{[\![c_{\mathbf{A}} \min\{\mathbf{A}\}]\!], \dots, [\![c_{\mathbf{A}} \max\{\mathbf{A}\}]\!]\}$, with $c_{\mathbf{A}}$ the companding coefficient ($c_{\mathbf{A}} > 0$) that is designed according to precision requirements for the final result (equivalently for $\tilde{b}_{m,n}$). These values are then used in the core of the calculation, which, for each output element, is a vector inner product with $L$ multiply-accumulate (MAC) operations:

$$\tilde{r}_{m,n} = \sum_{l=0}^{L-1} \tilde{a}_{m,l} \tilde{b}_{l,n}. \tag{4}$$

The final result is recovered after reverse companding:

$$\hat{r}_{m,n} = (c_{\mathbf{A}} c_{\mathbf{B}})^{-1} \tilde{r}_{m,n}. \tag{5}$$

The selection of $c_\mathbf{A}$ and $c_\mathbf{B}$, as well as the control of the overall process, is discussed in Section 5.

Despite the distortion $\frac{1}{L^2}\|\widehat{\mathbf{R}} - \mathbf{R}\|_F$ due to quantization, using (4) for the actual computation of $\tilde{r}_{m,n}$ would not lead to any acceleration in a programmable processor; however, once the inputs have been companded and rounded, we can create reduced-size input blocks by packing multiple inputs together via two different methods. Both methods aim for accelerated processing.

*B. Symmetric Packing*

In the first method, the packing process creates two blocks $\overline{\mathbf{A}}$ and $\overline{\mathbf{B}}$ with $L \times \frac{L}{W}$ and $\frac{L}{W} \times L$ coefficients (respectively) given by ($\forall m, n: 0 \le m, n < L, \overline{m} = \lfloor \frac{m}{W} \rfloor, \overline{n} = \lfloor \frac{n}{W} \rfloor$):

$$\overline{a}_{m,\overline{n}} = \sum_{i=0}^{W-1} z^i \tilde{a}_{m, W\overline{n}+i}, \tag{6}$$

$$\overline{b}_{\overline{m},n} = \sum_{i=0}^{W-1} z^{-i} \tilde{b}_{W\overline{m}+i, n}, \tag{7}$$

where $z$, $0 < z < 1$, is the utilized packing coefficient that controls the allocated space for each packed input within the floating point representation (details on the setting of $z$ will be given in Section 4) and $W$ the number of packings performed. Notice that, unlike previous work on block-based packed processing [13][14][20], the elements of *both* $\widetilde{\mathbf{A}}$ and $\widetilde{\mathbf{B}}$ are packed along input rows and columns (respectively). Due to the block major-format reordering, (6) and (7) perform MAC operations in sequential memory elements, thereby allowing the use of SIMD instructions for accelerated processing.

Processing occurs using the packed data, i.e. $\overline{\mathbf{R}} = \overline{\mathbf{A}}\overline{\mathbf{B}}$:

$$\forall m, n: \overline{r}_{m,n} = \sum_{j=0}^{L/W-1} \overline{a}_{m,j} \overline{b}_{j,n} = \sum_{j=0}^{L/W-1} \left( \sum_{i=0}^{W-1} \tilde{a}_{m, Wj+i} \tilde{b}_{Wj+i, n} + \sum_{i=0}^{W-1} \sum_{\substack{h=0 \\ h \ne i}}^{W-1} z^{i-h} \tilde{a}_{m, Wj+i} \tilde{b}_{Wj+h, n} \right). \tag{8}$$

The packed result of (8) contains the required output as well as $(W^2 - W)$ "side" outputs: $o_{i,h} = z^{i-h} \tilde{a}_{m, Wj+i} \tilde{b}_{Wj+h, n}$, $\forall i, h: 0 \le i, h < W \,\&\, i \ne h$. Notice that (8) is performed in the function's native representation. As such, any high-performance $L \times \frac{L}{W}$ by $\frac{L}{W} \times L$ subblock code for `sGEMM` or `dGEMM` can be used for (8), as indicated in the subblock processing of Figure 1(b). Due to companding and packing, (8) performs $W$ times the operations of conventional SIMD-based matrix multiplication; we term this approach as *turbo SIMD*.

Following the completion of the processing, unpacking of the results can be performed by ($\forall m, n$):

$$u_{m,n} = [\![\overline{r}_{m,n}]\!], \tag{9}$$

$$\tilde{r}_{m,n} = u_{m,n} - \left( z^{-1} [\![z u_{m,n}]\!] \right). \tag{10}$$

The unpacking process extracts the useful result from the packed output by: *(i)* the rounding operation to remove the first unneeded set of results, $o_{i,h}$ with $i > h$, of (8); *(ii)* removing the second unneeded set of results, $o_{i,h}$ with $i < h$, of (8) by (10). Reverse companding can be applied to each $\tilde{r}_{m,n}$ via (5).

### C. Asymmetric Packing

In the second method, only one input block is packed; either $\widetilde{\mathbf{A}}$ or $\widetilde{\mathbf{B}}$ can be packed – this selection does not affect the performance in the case of GEMM as both blocks have been reordered in block major format. This is similar to the packing proposed in previous work [14][20]. Assuming $\widetilde{\mathbf{A}}$ is chosen, once the inputs have been companded and rounded by (3), the packing process creates block $\overline{\mathbf{A}}$ with $\frac{L}{W} \times L$ coefficients given by ($\forall m,n: 0 \leq m,n < L$, $\overline{m} = \left\lfloor \frac{m}{W} \right\rfloor$):

$$\bar{a}_{\overline{m},n} = \sum_{i=0}^{W-1} z^i \tilde{a}_{W\overline{m}+i,n}, \tag{11}$$

where $z$, $0 < z < 1$, is the utilized packing coefficient. Notice that, unlike (6), the packing of (11) operates along the columns of $\widetilde{\mathbf{A}}$ in order to pack $W$ rows together. This means that, in order to use SIMD instructions for accelerated computation of (11), we can group consecutive elements of each row and apply each MAC operation of (11) to them with one SIMD instruction. Once (11) is completed, processing occurs via $\overline{\mathbf{R}} = \overline{\mathbf{A}}\widetilde{\mathbf{B}}$ ($\forall m,n: 0 \leq m,n < L$, $\overline{m} = \left\lfloor \frac{m}{W} \right\rfloor$):

$$\begin{aligned} \bar{r}_{\overline{m},n} &= \sum_{j=0}^{L-1} \bar{a}_{\overline{m},j} \tilde{b}_{j,n} \\ &= \sum_{j=0}^{L-1} \sum_{i=0}^{W-1} z^i \tilde{a}_{W\overline{m}+i,j} \tilde{b}_{j,n}. \end{aligned} \tag{12}$$

The packed result of (12) contains the output of groups of $W$ rows packed together. Since the processing is performed in the function's native representation, any high-performance $\frac{L}{W} \times L$ by $L \times L$ subblock code for `sGEMM` or `dGEMM` can be used for (12). Following the completion of the processing, unpacking of the results can be performed by the following iterative process [14] ($\forall m,n: 0 \leq m,n < L$, $\overline{m} = \left\lfloor \frac{m}{W} \right\rfloor$):

$$\tilde{r}_{\overline{m},n} = [\![\bar{r}_{\overline{m},n}]\!], \tag{13}$$

$$\forall i \in \{1, \dots, W-1\}: \left[ \bar{r}_{\overline{m}+i,n} = z^{-1}\left( \bar{r}_{\overline{m}+i-1,n} - \tilde{r}_{\overline{m}+i-1,n} \right), \tilde{r}_{\overline{m}+i,n} = [\![\bar{r}_{\overline{m}+i,n}]\!] \right]. \tag{14}$$

Reverse companding can be applied to each $\tilde{r}_{m,n}$ via (5).

### D. Discussion

*Remark 1 (Integer Processing and the Concept of Packed Processing)*: Following the companding stage, matrix multiplication could be constructed using 16-bit integer representations (16-bit integer SIMD instructions exist for all mainstream processors). However, integer conversion to and from floating point leads to significant overhead since it cannot be performed with SIMD instructions. Furthermore, unlike floating point representations where the maximum packed value can be flexible with graceful degradation in the results, a strict limit is set on the quantized values in integers in order to avoid overflow.

A conceptual illustration of the packed representation of (8) and (12) and how the floating-point representation noise affects the packed results is given in Figure 2. Symmetric packing is better protected from the numerical representation noise, because the "lower" side result (multiplied by $z = 0.0001$) is not used; this is despite the fact that noise is amplified in this representation as the "higher" side result (multiplied by $z^{-1} = 10000$) takes the number further away from the high-precision region around zero [21]. This representation noise creates the notion of *computational capacity* in our approach: for given quantization distortion, there is a limit on the throughput increase achieved via increased packing (i.e. increased values for $W$, surpassing $W = 2$ shown in Figure 2), beyond which the distortion stemming from the floating point computation surpasses the quantization distortion[3]. The interdependency between throughput and distortion and the notion of computational capacity make our approach a *computation channel* for generic matrix multiplication.    □

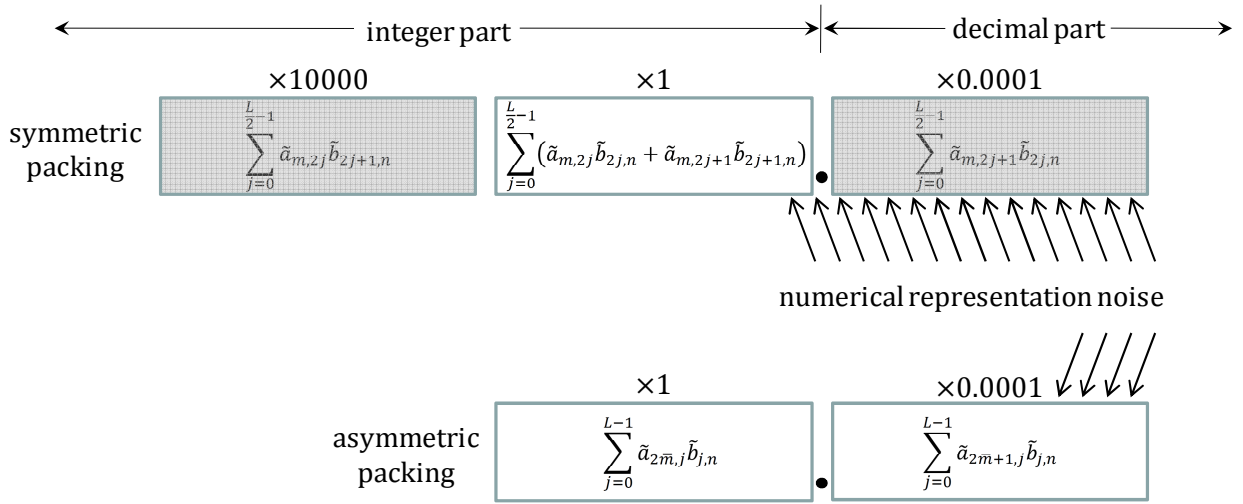

**Figure 2.** Conceptual example of $W = 2$ packings with $z = 0.0001$. **Top:** result of (8); **bottom:** result of (12). Shaded blocks contain side results that are produced during the packed processing but not used in GEMM.

*Remark 2 (Packing Tradeoff)*: The symmetric packing approach of Subsection 3.B produces $L \times L$ outputs, each requiring $L/W$ MAC operations. It requires $2L^2 W^{-1} b_{\text{repr}}$ bytes for storage of the packed input data, with $b_{\text{repr}}$ defined as for (1). On the other hand, the asymmetric packing of Subsection 3.C produces $\frac{L}{W} \times L$ outputs (that are then unpacked to the final $L \times L$ outputs), each requiring $L$ MAC operations. It requires $L^2(1 + W^{-1})b_{\text{repr}}$ bytes for storage of the input data. Evidently, both approaches have the same complexity in terms of MAC operations. They are both memory efficient in comparison to the conventional approach that requires $2L^2 b_{\text{repr}}$ bytes of memory for the input data. Between them, the symmetric packing requires the least amount of memory for the computation and, when the numerical representation noise is taken under consideration, it leads to better precision in comparison to symmetric packing as indicated previously. □

---

[3] Or, equivalently: decreasing the quantization distortion with the proposed packing (by increasing the companders) leads to increased noise stemming from the floating-point computations, as more space is needed to pack the quantized inputs and results (i.e. $z$ in Figure 2 becomes smaller); thus, for each packing (i.e. acceleration) $W$, there is a limit on the quantization accuracy, beyond which the distortion stemming from floating-point computation surpasses the quantization distortion.

## 4. GEMM AS A COMPUTATION CHANNEL – NUMERICAL REPRESENTATION NOISE, QUANTIZATION NOISE, AND COMBINED NOISE MODEL

*A. Summary of Known Results on Operational Packing*

The accuracy of the numerical representation sets limits on error-free packed processing of integers, which have been established within the tight packing theory [14][20].

*Proposition 1 (Error-free Packed Processing [14]):* Error-free unpacking of $W_{\text{ef}}$ signed integers by (9) and (10), or (13) and (14), after performing (8) or (12), respectively, requires:

$$z \leq \frac{1}{2R_{\max} + u_{\text{safe}}} \tag{15}$$

$$W_{\text{ef}} \leq \lfloor \log_z[(2R_{\max} + 1)u_{\text{sys}}] + 1 \rfloor, \tag{16}$$

with: $u_{\text{sys}}$ the relative precision of the computer hardware used for implementation [14], $u_{\text{safe}}$ a positive integer to avoid erroneous unpacking under marginal conditions [14] (in practice one can set $u_{\text{safe}} = 50$ to cover all possible scenarios with no practical loss in the packing capability) and, for the proposed companding process of (3), the maximum amplitude of the elements of the packed matrix $\overline{\mathbf{R}}$, $R_{\max}$, given by:

$$R_{\max} = \llbracket c_\mathbf{A} c_\mathbf{B} \times L \times \max\{|\min\{\mathbf{A}\}|, \max\{\mathbf{A}\}\} \times \max\{|\min\{\mathbf{B}\}|, \max\{\mathbf{B}\}\}\rrbracket. \tag{17}$$

*Proof:* See [14]. A simple algorithm to calculate $z$ and $W_{\text{ef}}$ for any processor is provided in [14][20].

The question we pose here is: *Can we go beyond the limits of error-free packed processing of Proposition 1 under the proposed throughput/distortion framework?* If this would be possible, we could approach the computational capacity of a given system under a specified distortion constraint. Exceeding the packing limitations was speculated by previous work [20], but it has never been investigated systematically until now.

There are two ways this may be possible. Firstly, one can attempt to increase $z$ beyond the limit of (15) in order to "squeeze in" more data in the floating-point representation. In such a case, the output results from the quantized processing of (8) or (12) may "invade" each other (i.e. the blocks of Figure 2 start to overlap) causing catastrophic error during unpacking [14][20]. Because of the severity of the errors caused, this is clearly an undesired option. As an alternative, one can attempt to utilize values for packing beyond $W_{\text{ef}}$ that is the limit set by (16). If one would apply such increased packing, distortion will *gradually* be introduced in the unpacked results of (10) and (13), (14). This may be acceptable since quantization already introduces approximation. To this end, by modifying $R_{\max}$ for every $W$, $W > 1$, one can systematically investigate the tradeoff between *quantization-induced* error and *representation-induced* error in order to approach the system's computational capacity: high values for $R_{\max}$ reduce the quantization error [since $c_\mathbf{A}$ and $c_\mathbf{B}$ can be increased in (17)] but may lead to significant representation-induced error if the bound of (16) is violated (i.e. if we use $W > W_{\text{ef}}$ for the selected $R_{\max}$ value); low values for $R_{\max}$ cause the reverse effect. Thus, via the use of companding and packing, we must identify the value of $R_{\max}$ for each packing $W$ leading to optimal throughput/distortion performance.

### B. *Noise of Packed Matrix Multiplication in Floating-point Representations*

In order to collect statistics from the representation-induced error under packed processing, we use integer samples for **A** and **B** with $L = 288$ and we set $c_\mathbf{A} = c_\mathbf{B} = 1$ in all these experiments, i.e. no loss is caused from companding and rounding. The experiments are performed with $\max\{\mathbf{A}\} = -\min\{\mathbf{A}\} = 22$ and $\max\{\mathbf{B}\} \in \{1,2,\ldots,63\}, \min\{\mathbf{B}\} = -\max\{\mathbf{B}\}$ using five random instantiations for **A** and **B** for each combination of maximum input values[4]. Using (17), these ranges and subblock size lead to output range $R_{\max} \in \{6336, 12672, \ldots, 399168\}$, which encompasses the range where $W_{\text{ef}} \in \{2,3,4\}$ is obtained in single or double-precision floating point representations. For each instantiation of each combination of $R_{\max}$ and $W$, we measure the mean error and the root mean squared error (RMSE) of each packed matrix multiplication after unpacking by:

$$\forall \max\{\mathbf{B}\}, \max\{\mathbf{A}\}, W\colon m(R_{\max}, W) = \tfrac{1}{L^2}\sum_{m=0}^{L-1}\sum_{n=0}^{L-1}(\hat{r}_{m,n} - r_{m,n}), \quad s(R_{\max}, W) = \tfrac{1}{L^2}\|\hat{\mathbf{R}} - \mathbf{R}\|_F \quad (18)$$

respectively, with: $W \in \{2,3,4\}$ packings, $z$ set for each $W$ according to (15) and **R** computed with $W = 1$ (conventional computation) under the same data type (single or double-precision floating point representation). When $m(R_{\max}, W) \cong 0$ (near-zero bias), $s(R_{\max}, W)$ approaches the sample standard deviation of the error. Furthermore, given that there is no quantization noise in this experiment, any mismatch in the results stems solely from the imprecision caused by the numerical representation. Specifically, depending on the value of $W$ and the input dynamic range, packed processing may not induce any error [14][20]. However, once $W > W_{\text{ef}}$ of (16) for the setting of $R_{\max}$ of a particular measurement, the output will contain numerical errors.

---

[4] The ranges for **A** and **B** were chosen so as to produce $R_{\max}$ values within the range required to cover the numerical representation limits for each number of packings $W$.

Figure 3(a) shows the relationship of RMSE with $R_{\max}$ for single and double-precision floating-point representations in an Intel Core 2 Duo P8800 processor (with $W = 2$ and $W = 4$, respectively). Figure 3(b) presents the average error for each case. The results of double precision with $W = 4$ are not displayed beyond $R_{\max} > 120000$ as they exceed acceptable limits: despite the increased accuracy of double-precision, four packings occupy significantly more space in the numerical representation; this makes the packed results of this case significantly more prone to numerical representation noise than the ones with two packings in single precision (see also Remark 1). This experiment indicates that, depending on the numerical representation (single or double-precision) and $W$, one can select a range of $R_{\max}$ values and safely assume near-zero average error $m(R_{\max}, W)$, i.e. no systematic error. In such a case, the sample standard deviation of the representation-induced error is monotonically increasing with $R_{\max}$ as shown in Figure 3(a).

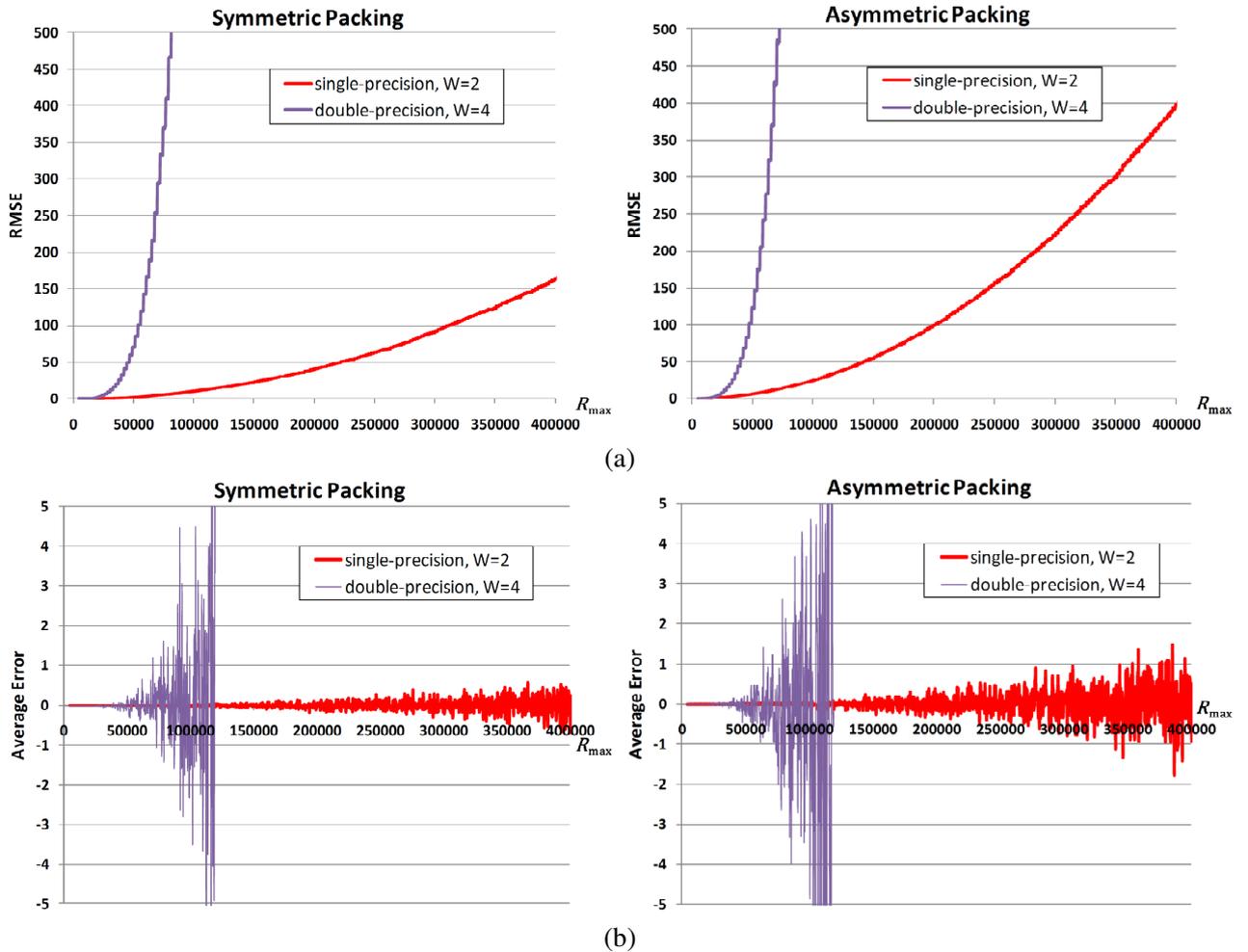

**Figure 3.** Error measurements for matrix multiplication of integer inputs ($c_\mathbf{A} = c_\mathbf{B} = 1$) within single-precision and double-precision representation: **(a)** root mean squared error; **(b)** average error.

Since each value of $m(R_{\max}, W)$ and $s(R_{\max}, W)$ reported in Figure 3 corresponds to hundreds of thousands of independent inner-product calculations [via (8) or (12)], the results of Figure 3 are a good approximation of the ensemble statistics. Thus, they are independent of the actual data being used for the matrix multiplication and stem solely from the numerical representation limitations and the utilized packing.

Interestingly, for symmetric packing in single-precision floating point representation with $W = 2$, we can use up to $R_{\max} = 320000$ and obtain: $\frac{m(R_{\max}, 2)}{R_{\max}} < 0.0001\%$ and $\frac{s(R_{\max}, 2)}{R_{\max}} < 0.04\%$. This indicates that one can utilize $R_{\max}$ values that are up to *an order of magnitude higher* than the maximum range of 16-bit signed integer representations (where $R_{\max} < 32768$ is imposed in order to avoid overflow) with very small relative error induced by the representation. Thus, the proposed packed processing in floating-point representations allows for significantly increased quantization range at the cost of gradually increased representation-induced noise [i.e. increased noise standard deviation $s(R_{\max}, W)$ for increased $R_{\max}$]. This noise is significantly smaller for the case of symmetric packing, as shown by the comparison between the two graphs of Figure 3(a). This occurs because, as mentioned in Remark 1, this packing does not use outputs $o_{i,h}$ of (8) with $i > h$, which are the outputs affected the most by the representation-induced noise. This part is indeed used in the asymmetric packing that actually extracts all outputs of (12) via the iterative unpacking of (13) and (14).

### C. Quantization Noise Model

We now present a statistical model of the quantization noise introduced via companding. Inputs $a_{m,n}$ and $b_{m,n}$ are modeled as zero-mean independent, identically distributed (iid), random variables (RVs) $\alpha \sim P_\alpha(\sigma_\alpha)$, $\beta \sim P_\beta(\sigma_\beta)$, respectively. Quantization noise in $\tilde{a}_{m,n}$, $\tilde{b}_{m,n}$ is modeled as additive, zero-mean, iid white, RVs $\nu_\alpha \sim P_{\nu_\alpha}(\sigma_{\nu_\alpha})$, $\nu_\beta \sim P_{\nu_\beta}(\sigma_{\nu_\beta})$ with $\sigma_{\nu_\alpha} = \frac{1}{c_A\sqrt{12}}$, $\sigma_{\nu_\beta} = \frac{1}{c_B\sqrt{12}}$, i.e. the standard deviation of the noise per matrix is scaled according to the companding applied. Finally, the output results $\hat{a}_{m,n}$ are modeled by zero-mean iid RVs $\hat{\rho} \sim P_{\hat{\rho}}(\sigma_{\hat{\rho}})$.

*Proposition 2 (Quantization Noise Power)*: Under iid inputs modeled by $\alpha \sim P_\alpha(\sigma_\alpha)$, $\beta \sim P_\beta(\sigma_\beta)$ and iid quantization noise modeled by $\nu_\alpha \sim P_{\nu_\alpha}(\sigma_{\nu_\alpha})$, $\nu_\beta \sim P_{\nu_\beta}(\sigma_{\nu_\beta})$, the expected noise power of the output results of the matrix multiplication under error-free unpacking is ($0 \leq m, n < L$):

$$E\{(\rho_{m,n} - \hat{\rho}_{m,n})^2\} = L\left[(\sigma_\alpha \sigma_{\nu_\beta})^2 + (\sigma_\beta \sigma_{\nu_\alpha})^2 + (\sigma_{\nu_\alpha} \sigma_{\nu_\beta})^2\right]. \tag{19}$$

*Proof:* See Appendix.

The expected power of the (error-free) output **R** is $L(\sigma_\alpha \sigma_\beta)^2$. Hence, the expected signal to noise ratio (SNR) of $\hat{\mathbf{R}}$ versus **R** is:

$$E\{S_{\text{subblock}}\} = 10\log_{10} \frac{L(\sigma_\alpha \sigma_\beta)^2}{E\{(\rho_{m,n} - \hat{\rho}_{m,n})^2\}} \tag{20}$$

Notice that, if the statistics of the input matrices are known (i.e. assuming known or estimated $P_\alpha$ and $P_\beta$), (19) and (20) are parameterized solely by $c_\mathbf{A}$ and $c_\mathbf{B}$.

### D. Combined Noise Model

Since the quantization noise of Proposition 2 and the representation-induced noise of Subsection 4.B stem from physically independent processes, we can assume they are statistically independent. The expected noise power of the output results is then:

$$D_W(R_{\max}, c_\mathbf{A}, c_\mathbf{B}) = L\left[(\sigma_\alpha \sigma_{\nu_\beta})^2 + (\sigma_\beta \sigma_{\nu_\alpha})^2 + (\sigma_{\nu_\alpha}\sigma_{\nu_\beta})^2\right] + \left[\frac{1}{c_\mathbf{A} c_\mathbf{B}} s(R_{\max}, W)\right]^2 \quad (21)$$

where $s(R_{\max}, W)$ is the experimentally-measured RMSE of the representation-induced noise for values of $R_{\max}$ and $W$ that correspond to $m(R_{\max}, W) \cong 0$ [examples for $W = 2$ and $W = 4$ are given in Figure 3(a)], and factor $(c_\mathbf{A} c_\mathbf{B})^{-2}$ maps the representation-induced noise from the quantization index domain to the output value domain [via the reverse companding of (5)].

For a given choice of packing $W$, $W > 1$, we can express the intuitive trade-off between quantization and representation-induced noise by combining (21) and (17): increasing $R_{\max}$ leads to increased values for $c_\mathbf{A}$ and $c_\mathbf{B}$ and therefore reduced quantization noise variances $\sigma_{\nu_\alpha}$ and $\sigma_{\nu_\beta}$ and reduced quantization noise power from Proposition 2; however, as shown in Figure 3(a), the representation-induced noise $s(R_{\max}, W)$ increases monotonically with $R_{\max}$. Hence, $D_W$, as expressed by (21), becomes the mechanism for adjusting the desired throughput and distortion of GEMM according to user-specified constraints for: *(i)* percentile throughput increase against the conventional (full precision) computation $\mathbf{R} = \mathbf{AB}$ and *(ii)* the SNR (distortion) of $\widehat{\mathbf{R}}$ versus $\mathbf{R}$.

## 5. DISTORTION-CONTROLLED THROUGHPUT SCALING OF SUBBLOCK MULTIPLICATION

When attempting to accelerate GEMM with the proposed approach, it is imperative to minimize $D_W$ for each $L \times L$ subblock multiplication of each $L \times L$ inner-kernel calculation of the $M \times K$ by $K \times N$ matrix multiplication [e.g. $\forall l: \mathbf{A}_{2,l}\mathbf{B}_{l,1}$ of $\mathbf{R}_{2,1}$ of Figure 1 and (2)] via the optimal use of companding and packing. Once the optimal configuration and minimum distortion is established for each packing $W$ of each subblock, the configuration for companding and packing can be decided for the overall matrix multiplication.

### A. Theoretical Calculation of Optimal Companders and Experimental Validation

We provide the general form of admissible companders for each $L \times L$ subblock under iid inputs.

*Proposition 3 (General Form of Companders):* For quantized packed subblock multiplication $\widetilde{\mathbf{R}} = \widetilde{\mathbf{A}}\widetilde{\mathbf{B}}$ of zero-mean iid inputs within $[\min\{\mathbf{A}\}, \max\{\mathbf{A}\}]$ and $[\min\{\mathbf{B}\}, \max\{\mathbf{B}\}]$ modeled by $\alpha \sim P_\alpha(\sigma_\alpha)$, $\beta \sim P_\beta(\sigma_\beta)$, respectively, and quantization noise modeled by $v_\alpha \sim P_{v_\alpha}(\sigma_{v_\alpha})$, $v_\beta \sim P_{v_\beta}(\sigma_{v_\beta})$, the companders achieving expected SNR of $E\{S_{\text{subblock}}\}$dB against $\mathbf{R} = \mathbf{AB}$ calculated with floating-point precision are:

$$c_{\mathbf{A}} = \frac{1}{\sqrt{2}\sigma_\alpha c_{\text{tot}}}\sqrt{D_{\text{QvR}} \pm \sqrt{D_{\text{QvR}}^2 - 4\sigma_\alpha^2 \sigma_\beta^2 c_{\text{tot}}^2}}, \qquad c_{\mathbf{B}} = \frac{\sqrt{2}\sigma_\alpha}{\sqrt{D_{\text{QvR}} \pm \sqrt{D_{\text{QvR}}^2 - 4\sigma_\alpha^2 \sigma_\beta^2 c_{\text{tot}}^2}}} \qquad (22)$$

with $D_{\text{QvR}}$ expressing the *quantization-versus-representation induced distortion*, given by:

$$D_{\text{QvR}} = \frac{12\sigma_\alpha^2 \sigma_\beta^2}{10^{0.1 E\{S_{\text{subblock}}\}}} - \frac{[12 s(R_{\max}, W)]^2 + 1}{12 c_{\text{tot}}^2} \qquad (23)$$

$$c_{\text{tot}} = \frac{L \times \max\{|\min\{\mathbf{A}\}|, \max\{\mathbf{A}\}\} \times \max\{|\min\{\mathbf{B}\}|, \max\{\mathbf{B}\}\}}{R_{\max}} \qquad (24)$$

and
$$D_{\text{QvR}} \geq 2\sigma_\alpha \sigma_\beta c_{\text{tot}}. \qquad (25)$$

*Proof:* See Appendix.

Given the input statistics, Proposition 3 demonstrates that various pairs of companders provide for $E\{S_{\text{subblock}}\}$dB as long as they lead to $R_{\max}$ [via (17)] that satisfies (25). This complicates the selection process for the operational parameters since, per $L \times L$ subblock of the matrix multiplication, one must select:

*(i)* the number of packings used ($W$),

*(ii)* the desired value of $R_{\max}$ in order to find $s(R_{\max}, W)$ for this choice of packed processing via Figure 4(a),

*(iii)* the desired value for $E\{S_{\text{subblock}}\}$ for the particular subblock [leading to $D_{\text{QvR}}$ that satisfies (25)], and

*(iv)* the specific set of companders from the permissible options of (22).

Fortunately, in the following we show that, for each subblock and each choice of packing, $W \in \{2,3,4\}$, there exists a *unique* value for each of: $R_{\max}$, $c_{\mathbf{A}}$, and $c_{\mathbf{B}}$, which maximizes the expected SNR $E\{S_{\text{subblock}}\}$. An expression for the obtained (maximum) value of $E\{S_{\text{subblock}}\}$ for each $W$ is also provided. This facilitates the parameter selection to a great extent as there is a unique (optimal) parameter configuration for each packing $W$ of each subblock.

*Proposition 4 (Minimum-error Companders):* Under the settings of Proposition 3 for subblock multiplication, for each packing $W > 1$, the companders providing the maximum expected SNR of $E\{S_{\text{subblock}}^*\}$dB are:

$$c_{\mathbf{A}}^* = \sqrt{\frac{\sigma_\beta}{\sigma_\alpha c_{\text{tot}}}}, \qquad c_{\mathbf{B}}^* = \sqrt{\frac{\sigma_\alpha}{\sigma_\beta c_{\text{tot}}}} \qquad (26)$$

with $\{R_{\max}^*, E\{S_{\text{subblock}}^*\}\}_W = \arg\max_{\forall R_{\max}} \left\{ E\{S_{\text{subblock}}\} = -10\log_{10}\left[\frac{[12 s(R_{\max}, W)]^2 + 1}{144\sigma_\alpha^2 \sigma_\beta^2 c_{\text{tot}}^2} + \frac{c_{\text{tot}}}{6\sigma_\alpha \sigma_\beta}\right] \right\}$ (27)

for every $W > 1$ and $c_{\text{tot}}$ given by (24).

*Proof:* See Appendix.

We demonstrate indicative experimental versus theoretical (model) curves of $E\{S_{\text{subblock}}\}$ in Figure 4. Per $R_{\max}$ value, the experimental curves are produced by measuring the average SNR, $S_{\text{subblock}}$, numerically from the output of multiple runs under the experimental settings of Subsection 4.B but this time using floating-point inputs instead of integers, and companders set via (26). For the theoretical calculation for each $R_{\max}$ value, we use the expression of (27) for $E\{S_{\text{subblock}}\}$ with $s(R_{\max}, W)$ taken from the results of Figure 3(a). The very good agreement between the theoretical and experimental results demonstrates the accuracy of the proposed SNR estimation model of (27). The unique maximum SNR observed in the results of Figure 4 shows that, under the conditions of Proposition 3, there is indeed a unique solution for companders per subblock (and per $W$) that minimizes the produced error, which is given by (27) of Proposition 4.

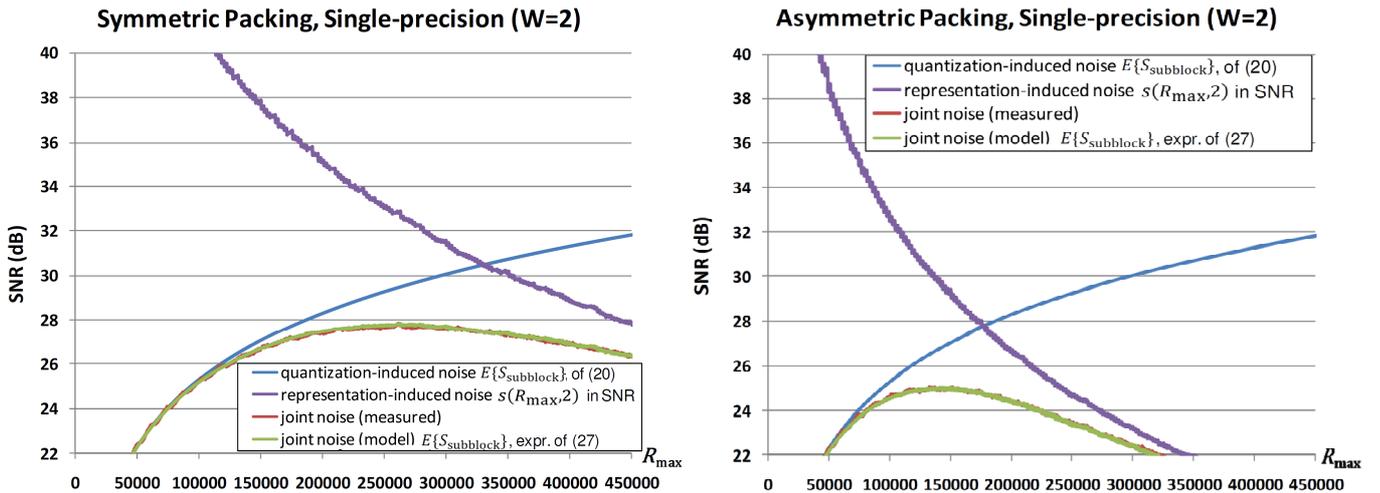

**Figure 4.** Indicative experiments for experimentally-obtained output SNR versus the expected SNR by Proposition 4 under the use of the optimal companders per $R_{\max}$ value.

Finally, the independence assumption between quantization-induced and representation-induced error was further validated by calculating the Pearson correlation coefficient, $\frac{E\{\varepsilon_{\text{quant}}\varepsilon_{\text{repr}}\}}{\sqrt{E\{\varepsilon_{\text{quant}}^2\}}\sqrt{E\{\varepsilon_{\text{repr}}^2\}}}$ with $\varepsilon_{\text{quant}}$ and $\varepsilon_{\text{repr}}$ the RVs of the error under solely quantization-induced and solely representation-induced noise (respectively[5]), from the ensemble of the experiments reported in Figure 4 and the corresponding experiments for double-precision representation. In both cases, the correlation coefficient was found to be smaller than $10^{-2}$. This fact, along with the fact that the quantization-induced and representation-induced noise terms stem from physically-separated independent processes and that the model of (27) agrees very well with the experimental results, is a strong indication that the two error terms are indeed uncorrelated and that the distortion expressed in (21) is accurate.

---

[5] Thus, $E\{\varepsilon_{\text{quant}}^2\}$ is given by (19) and $\sqrt{E\{\varepsilon_{\text{repr}}^2\}} = s(R_{\max}, W)$, which is given by (18).

*B.  Practical GEMM Configuration for Optimized Throughput/Distortion Processing*

Proposition 4 simplifies the optimum selection of operational settings for the overall matrix multiplication under throughput or distortion constraints. This is achieved by first computing min{**A**}, max{**A**}, min{**B**}, max{**B**}, $\sigma_\alpha$, $\sigma_\beta$ during the subblock data accesses for reordering to block-major format at runtime. Subsequently, for every $L \times L$ subblock multiplication of every $L \times L$ inner-kernel of the overall $(M \times K) \times (K \times N)$ matrix multiplication, $c_\mathbf{A}^*, c_\mathbf{B}^*, \{R_{\max}^*, E\{S_{\text{subblock}}^*\}\}$ are computed via Proposition 4 for every packing $W$, $W \in \{2,3,4\}$, at runtime. For each $W$, the expected percentile throughput scaling, $F_W$, versus the plain processing ($W = 1$) can be calculated by off-line experiments on the target platform (e.g. during the package compilation [5]), since it depends only on the subblock size and the implementation of the inner-kernel processing. These optimal parameters for each packing configuration $W$ of each subblock, along with $F_W$, are kept in a data structure in order to prune out the best possible combination of subblock multiplication options (at runtime) according to distortion or throughput constraints, as discussed in the next section. The only complex part of this process is obtaining the numerical maximum of (27) per subblock. However, Figure 4 shows that obtaining the exact solution for $R_{\max}^*$ is not of critical importance, since there is a wide range of $R_{\max}$ values attaining near maximum SNR per subblock. Hence, we pre-compute (offline) an approximation for all possible solutions of (27) once, by using a representative set of input block standard deviations, $\{\sigma_\alpha, \sigma_\beta\}_{\text{offline}}$, expected to occur at runtime. We keep the corresponding solutions in a data structure and, during the GEMM runtime operation, for each pair of subblocks $\mathbf{A}_{i,l}$ and $\mathbf{B}_{l,j}$ ($0 \leq l < \frac{K}{L}$), we select the solution with standard deviations $\sigma_\alpha$, $\sigma_\beta$ closest to the obtained ones (in the MSE sense), which requires minimal effort.

## 6.  THROUGHPUT/DISTORTION OPTIMIZATION OF INNER-KERNEL MULTIPLICATION

We can now perform throughput/distortion optimization that is controlled at the inner-kernel level of the overall matrix multiplication. Consider the example of the inner-kernel result $\mathbf{R}_{2,1}$ shown in Figure 1(a) and computed by (2). As elaborated in the previous section, for each individual subblock multiplication of (2), i.e. $\mathbf{A}_{2,l}\mathbf{B}_{l,1}$ with $0 \leq l < \frac{K}{L}$, we have precomputed parameters: $\forall W, l: \{R_{\max}^*, E\{S_{\text{subblock}}^*\}\}|_{\text{offline}}$ and $F_W$, and for each packing we can compute $c_\mathbf{A}^*, c_\mathbf{B}^*$ by (26) at runtime [and from that we can readjust $E\{S_{\text{subblock}}^*\}$ via the expression of (27) for additional accuracy]. The aim is to select (per subblock) the highest possible packing, $W$, and companders such that the inner-kernel, $\mathbf{R}_{2,1}$, is derived close to the computational capacity of the system, i.e. with:

*(i)* the highest percentile acceleration possible under an SNR constraint, or

*(ii)* the highest SNR possible under a percentile acceleration constraint.

Controlling the throughput/distortion optimization process at the inner-kernel level allows for flexibility within practical applications. For example, if matrix multiplication is used for transform decomposition, some inner-kernels of the resulting matrix corresponding to transform coefficients (e.g. frequency components) that are required at the smallest possible distortion can be computed with $W = 1$, i.e. with native floating point accuracy of sGEMM or dGEMM, while others can be accelerated via the use of $W = \{2,3,4\}$ and produce approximate results. The application must only specify the required (minimum) SNR $S_{\text{kernel}}(i,j)$ per $L \times L$ inner-kernel $\mathbf{R}_{i,j}$ ($0 \leq i < \frac{M}{L}, 0 \leq j < \frac{N}{L}$) against the result at the native precision of GEMM, or the required percentile acceleration, $F_{\text{kernel}}(i,j)$, in comparison to computing with $W = 1$ and the pruning process described in the following will derive the appropriate settings per subblock multiplication in order to meet this specification. For each inner kernel $\mathbf{R}_{i,j}$, $S_{\text{kernel}}(i,j)$ is converted to MSE by $D_{\text{kernel}}(i,j) = 10^{-0.1 S_{\text{kernel}}(i,j)} L \sum_{l=0}^{K/L-1} (\sigma_{\alpha,i,l} \sigma_{\beta,l,j})^2$, with $\sigma_{\alpha,i,l}$ and $\sigma_{\beta,l,j}$ the standard deviation of $\mathbf{A}_{i,l}$ and $\mathbf{B}_{l,j}$, respectively.

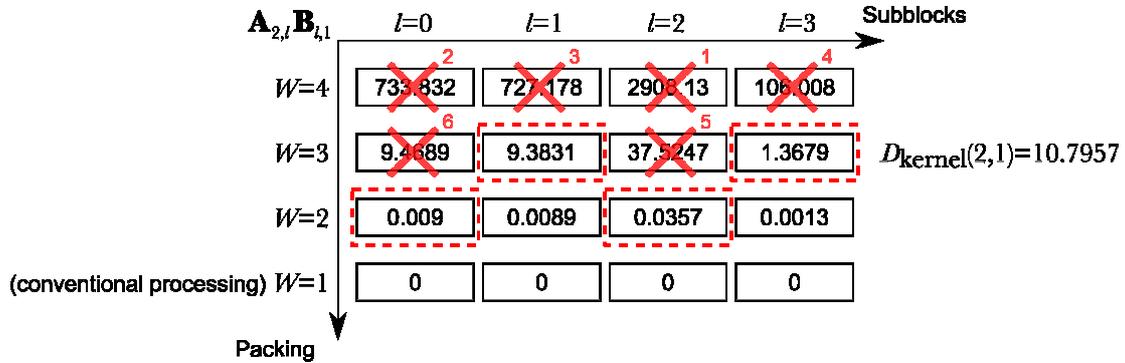

**Figure 5.** Example distortion (MSE) for each of the subblocks of (2) with $\frac{K}{L} = 4$ and pruning steps enumerated until $D_{\text{kernel}}(2,1) \leq 11.0$ is achieved. The dashed rectangles indicate the final selection for each subblock.

The utilized pruning is a top-down approach where, starting from the maximum acceleration, the selection is pruned by removing the outcome with the highest distortion, until the distortion constraint, $D_{\text{kernel}}(i,j)$, or the percentile throughput acceleration constraint, $F_{\text{kernel}}(i,j)$, is met. This is illustrated for distortion-constrained processing of inner-kernel $\mathbf{R}_{2,1}$ in the example of Figure 5, where we assume $\frac{K}{L} = 4$ and we set $S_{\text{kernel}}(2,1) \geq$ 64dB, which corresponded to $D_{\text{kernel}}(2,1) \leq 11.0$ under the utilized settings. For every calculation $\mathbf{A}_{2,l} \mathbf{B}_{l,1}$, $0 \leq l < 4$, the algorithm starts from $W = 4$ and successively removes the subblock result with the highest distortion (the removal steps are enumerated), until the distortion constraint is met. Thus, the resulting settings utilize the maximum packing possible (i.e. offer the maximum acceleration) under a distortion constraint.

## 7. EXPERIMENTAL RESULTS

We implemented the proposed framework in its entirety using streaming SIMD extensions (SSE3) in an Intel Core 2 Duo P8800 processor operating at $c_{\text{freq}} = 2.66\text{GHz}$ (Ubuntu Linux, single-threaded execution, `gcc4.4.1 -O3 -march=native -fomit-frame-pointer`, CPU throttling disabled to ensure maximum performance). In our experiments we selected: $W \in \{1,2\}$ for single precision floating point; $W \in \{1,2,3,4\}$ for double precision; and $L = 288$ as a representative inner-kernel size.

For the generic experiments of this section, we set input matrices **A** and **B** to contain uniformly distributed floating-point inputs selected from $[-\max_e, \max_e]$ within subblocks of $288 \times 288$, with $\max_e$ selected randomly for each subblock from the set $\{4.0, 5.0, 6.0, \ldots, 2048.0\}$. By disabling the processor throttling and running the proposed approach in maximum priority, we set various SNR requirements $\forall i, j: S_{\text{kernel}}(i,j)$ for each inner-kernel processing and obtained the results for $M = K = N$ shown in Figure 6. As an external comparison, we provide the performance of GEMM without the proposed approach (``sGEMM plain'', ``dGEMM plain'') as well as the performance of the state-of-the-art ATLAS [5] and GOTO packages [22]. With the selected input dimensions, all packages avoid "cleanup" code for the borders of the matrix multiplication. The full source code of our proposal is available online for inspection and testing [17].

The results reported in Figure 6 show that the proposed approach provides for processing throughput that changes according to the required SNR value and it can in fact exceed 130% and 175% of the peak performance for single-precision and double-precision representation, respectively. Although this is theoretically possible due to the utilized companding and packing, the results of Figure 6 show that *exceeding 100% of peak performance is indeed possible in practice*. Notice that the companders for each inner kernel are found at runtime and the performance reported in Figure 6 includes the entire process and the control framework described in Sections 5 and 6. Finally, we validated that, even when the SNR setting leads to $W = 1$ for all inner-kernel processing, no loss in performance is observed against the conventional ``sGEMM plain'' and ``dGEMM plain'' approaches.

Beyond distortion-controlled execution, we present an example of throughput-controlled acceleration in Figure 7 for the case of `sGEMM`, where $W \in \{1,2\}$. The 11 experimental points reported in the figure were obtained for matrix multiplication of size $M = N = K = 4032$ by increasing the percentage of accelerated inner-kernel blocks in steps of 10%: the (leftmost) maximum-SNR point (infinity – out of the graph's range) corresponds to $F_{\text{kernel}} = 0\%$ of inner-kernel blocks accelerated (i.e. $W = 1$ for all – conventional computation), while the (rightmost) lowest-SNR point (27.8dB for symmetric packing and 23.9dB for asymmetric packing) corresponds to $F_{\text{kernel}} = 100\%$ of inner-kernel blocks accelerated (i.e. $W = 2$ for all). GOTO's throughput was 19.7GFLOPS while ATLAS achieved 18.0GFLOPS for this example. Evidently, the symmetric packing provides for lower distortion along the operational points. This agrees with the error measurements of Figure 3 for symmetric and asymmetric packing and it occurs for the reason explained in Remark 1 and quantified in Section 4.B.

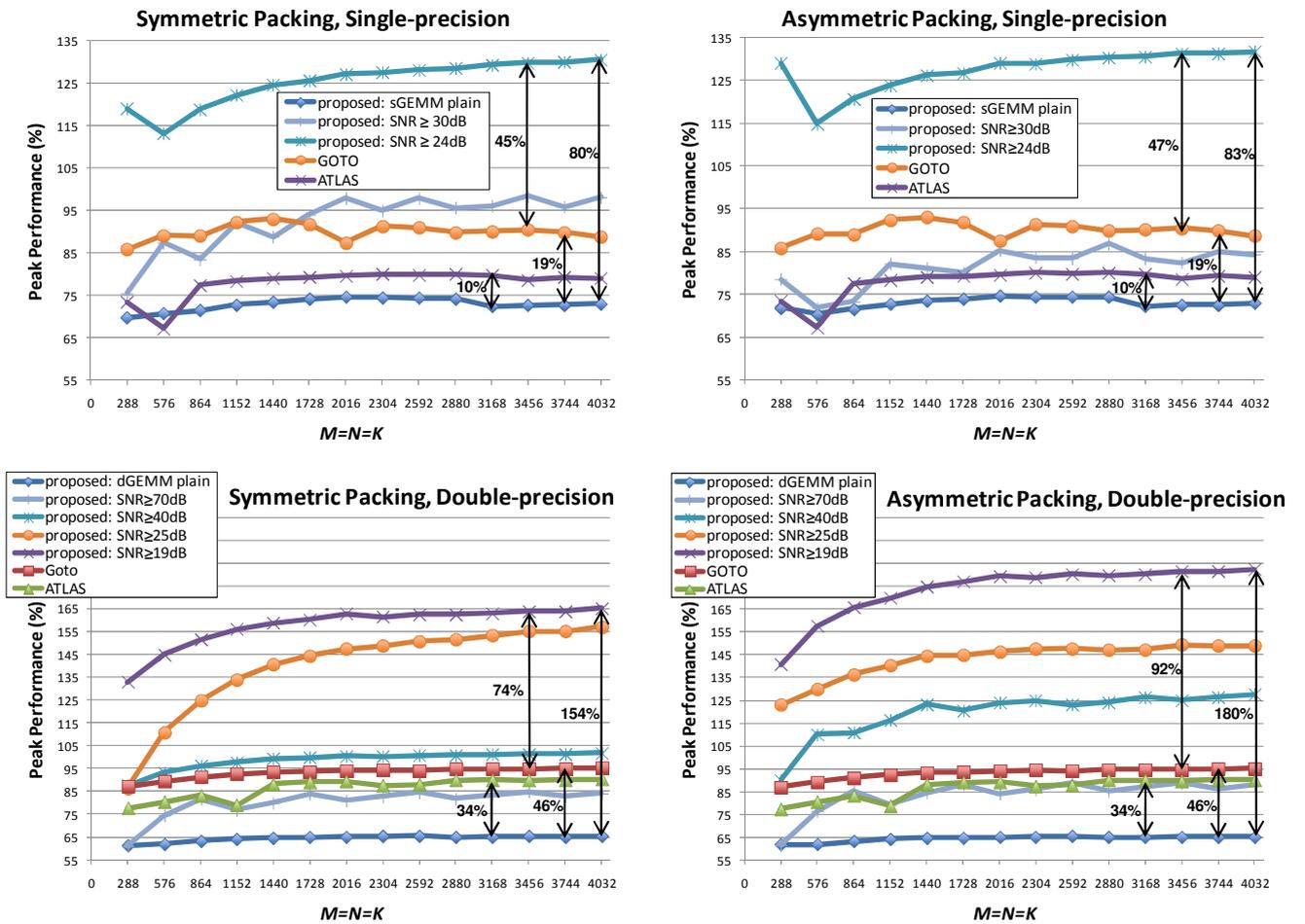

**Figure 6.** Percentage of peak performance of sGEMM and dGEMM (higher is better) under different SNR settings $\forall i,j: S_{\text{kernel}}(i,j)$ for the inner kernel processing; 100% of peak performance [22] corresponds to $8 \times c_{\text{freq}} = 21.28$GFLOPS (Giga floating-point operations per second) for sGEMM and $4 \times c_{\text{freq}} = 10.64$GFLOPS for dGEMM.

Overall, due to the highly-optimized nature of GOTO and ATLAS, our "sGEMM plain" subblock code for matrix multiplication (i.e. the conventional full-precision matrix multiplication core implemented for the purposes of this paper [17]) is 19% less efficient than GOTO and approximately 10% less efficient than ATLAS. In double precision, the equivalent loss in performance is 46% and 34%, respectively. This indicates that there is room for further improvement: if our approach were to be deployed with GOTO's (or any other) subblock code, landmark performance of beyond *150% of peak performance in single precision* and beyond *200%* of *peak performance in double-precision* representation could be achieved under throughput/distortion scaling.

The gains in processing throughput can be exchanged for fault detection and correction functionalities under error-generating operating systems or processors [12][18][19]. Alternatively, one can reduce the operating processor frequency (and operating voltage – via dynamic voltage scaling) and still obtain comparable performance to using "`sGEMM plain`" (or "`dGEMM plain`") at a higher processor frequency, albeit at higher distortion. This approach effectively translates the throughput/distortion scaling into power/distortion.

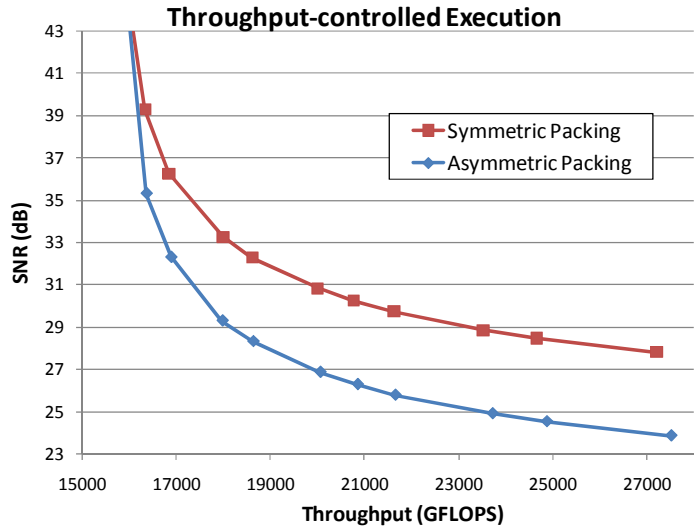

**Figure 7.** SNR vs. throughput in Giga floating-point operations per second for `sGEMM` under throughput control.

## 8. APPLICATIONS

The proposed approach can bring important benefits to high-performance signal processing systems when the precision of computation is not of critical importance (error-tolerant systems) or when the input data is intrinsically noisy. A conceptual application of disturbance cancellation under estimation uncertainty and power-distortion scalability with the proposed approach was already demonstrated in our recent work [16]. Here, we present two other applications of the proposed framework.

### A. *Accelerated Processing in State-of-the-art Face Recognition*

State-of-the-art techniques for robust image recognition systems derive feature matrices and use 2D decomposition schemes via matrix multiplication in order to match features between a new image and an existing database of images (e.g. for automatic identification of human faces [1]). Large-scale deployments of such systems run in high-performance workstations or in cloud computing infrastructures, such as Amazon's EC2. In such deployments, it is not uncommon to expect that thousands of training and recognition tasks should be computed with the highest-possible throughout/precision capability of each server in order to maximize the servers' or cloud utilization. Using the proposed approach, one can accelerate the real-time training and matching process. Specifically, the accelerated GEMM can be used for the image covariance matrix calculation and for the input image projection to the feature matrix [1]. In the following, we provide details of such a deployment for the prominent 2D PCA system of [1] under different SNR values for GEMM.

The algorithm consists of three stages: training, feature extraction and classification. The training stage uses a number of *training input images* and first calculates the image covariance scatter matrix from $J_{\text{set}}$ zero-mean input images, $\mathbf{A}_j$, by: $\mathbf{G}_J = \sum_{j=1}^{J_{\text{set}}} \mathbf{A}_j \mathbf{A}_j^{\text{T}}$. Based on this input training set, it then calculates the projection matrix comprising a series of projection axes (eigenvectors), $\mathbf{X} = [\mathbf{x}_1|...|\mathbf{x}_d]$, with $\mathbf{x}_i$, $1 \leq i \leq d$ the orthonormal eigenvectors of $\mathbf{G}_J$ corresponding to its $d$ largest eigenvalues [1]. Each training-set image is mapped to $\mathbf{X}$ via: $\mathbf{Y}_{\text{set},j} = \mathbf{A}_j \mathbf{X}$. For the feature extraction stage, each new input image $\mathbf{B}_i$ (*test image*) is mapped to $\mathbf{X}$ via: $\mathbf{Y}_{\text{test},i} = \mathbf{B}_i \mathbf{X}$. Finally, the classification stage determines for each test image $\mathbf{B}_i$ the test-set image $j^*$ with the smallest distance in their projections: $j^*_{\mathbf{B}_i} = \arg\min_{\forall j} \|\mathbf{Y}_{\text{test},i} - \mathbf{Y}_{\text{set},j}\|_F$. The complexity of 2D PCA is predominantly in the matrix multiplications required for the construction of $\mathbf{G}_J$ and the mapping during the feature extraction, i.e. $\mathbf{Y}_{\text{test},i}$, as the eigenvalue decomposition required for the creation of $\mathbf{X}$ is only performed once every $J_{\text{set}}$ training images and very fast algorithms exist for the quick estimation of $j^*_{\mathbf{B}_i}$ (e.g. matching error measures [23]).

To examine the effects of throughput/distortion within this application, we utilize the proposed single-precision GEMM design for the matrix multiplication operations of 2D PCA. The Yale database of images (http://www.face-rec.org/databases/) was used for our experiments and, following prior work [1], each image was cropped to $288 \times 288$ pixels (that includes the face portion) and the mean value was subtracted prior to processing. Results from performing all matrix multiplication operations with reduced SNR, $S_{\text{kernel}}(0,0)$, are presented in Table 1. Following [1], the first 5 images of each of the 15 persons in the database were used for the training set and the remaining 6 images per person were used as test images and we set $d = 10$. The table demonstrates that decreasing $S_{\text{kernel}}$ (for all GEMM subblocks) leads to increased processing throughput with the recognition accuracy remaining equal to the one obtained with the full-precision computation. In fact, for $S_{\text{kernel}} = 20$dB, we observed a slight increase in the recognition accuracy due to the quantization acting as a noise removal mechanism. We do not present results with values below $S_{\text{kernel}} = 20$dB, as 20dB allowed for $W = 2$ packings in all GEMM computations, which is the maximum acceleration possible under single precision. Importantly, we obtain the results of Table 1 without any application-specific tailoring of the computation; instead, only a simple adjustment of the distortion specification is required and the optimization framework of Section 6 selects the appropriate companding and packing parameters for each GEMM computation at runtime. To demonstrate the importance of throughput/distortion optimization in practice, "Proposed, ad-hoc" in Table 1 shows the results obtained with $W = 2$, $R_{\max} = 50000$, $c_A = c_B = 0.1$ (similar results were obtained with other ad-hoc variants); the severe drop in accuracy shows that ad-hoc parameter tuning does not suffice.

| Method | Recognition (%) | Throughput (GFLOPS) |
|---|---|---|
| GOTO [22] | 78.4 | 18.27 |
| Proposed: `sGEMM plain` | 78.4 | 14.85 |
| Proposed: $S_{\text{kernel}} = 50$dB | 78.4 | 16.11 |
| Proposed: $S_{\text{kernel}} = 20$dB | 78.8 | 25.31 |
| Proposed, ad-hoc | 44.0 | 25.31 |

**Table 1.** Recognition accuracy versus requested SNR for the matrix operations of 2D PCA and versus the obtained throughput for all matrix multiplication operations (higher throughput is better).

*B. Accelerated Supervised Training of Multi-layer Perceptron (MLP) based Learning System*

We examine the benefits of the proposed approach in a large deployment of a multi-layer perceptron-based learning system [3]. MLP-based learning uses the back-propagation algorithm (BPA) to train a neural network to create connections between input features and outputs. BPA is operating in bunch mode [3]: instead of using one training pattern at a time to update the weight matrices, the design uses $n_p > 1$ training patterns, which leads to matrix multiplications being used (and comprising the dominant part of the execution).

In order to derive test results, we utilized the Million-song dataset from Columbia University, available via the UCI Machine Learning Repository [15]. Our target was to predict the publication year of each song (between 1920-2010) based on the provided set of 90 features per song. MLP-based approaches are appropriate for such problems as there is no clear methodology to connect song features and publication year and the hope is that the learning algorithm will discover such connections automatically.

For our experiment, the bunch size was set to $n_p = 384$ and training was repeated in groups of 24576 songs randomly chosen from the training subset of the database. Validation was done on groups of 24576 songs from the validation subset of the database [15]. The only modification performed in the MLP implementation was the replacement of the GEMM implementation with: `sGEMM` from the GOTO library [22], `sGEMM plain`, and sGEMM with $F_{\text{kernel}} = 100\%$ (for all GEMM subblocks, corresponding to using $W = 2$ everywhere); here we investigate throughput-controlled acceleration. All experiments were executed on a high-performance server (2 quad-core Intel Xeon X5460 at 3.16GHz).

Table 2 reports summary results, showing that sGEMM with $F_{\text{kernel}} = 100\%$ packings (and companders selected via Proposition 4 at runtime based on $\sigma_\alpha$ and $\sigma_\beta$) achieves the same recognition accuracy as the conventional (full-precision) approaches. In this case, companding and rounding the inputs corresponds to quantizing to 13-15 bits. While it is well known that the BPA is robust to quantization, the quantization noise can increase the number of epochs required for training, as shown by Table 2. Importantly, sGEMM with $F_{\text{kernel}} = 100\%$ achieves 21% execution time reduction in comparison to sGEMM plain (i.e. approximately 4 days less). Despite this reduction, in this case sGEMM with $F_{\text{kernel}} = 100\%$ does not outperform GOTO due to its highly-optimized software implementation as compared to our own sGEMM plain design. However, a deployment of our approach using GOTO's subblock code would indeed benefit from the demonstrated execution time speedup.

| Method | Average number of epochs | Average recognition accuracy (correct/total) | Total time for GEMM computations (hours) |
|---|---|---|---|
| Proposed: sGEMM plain | 510240 | 14324/24576 (58%) | 474.6 |
| GOTO [22] | 510240 | 14324/24576 (58%) | 376.6 |
| Proposed: sGEMM $F_{\text{kernel}} = 100\%$ | 543460 | 14324/24576 (58%) | 375.5 |

**Table 2.** Summary results of MLP algorithm. Smaller time values are better.

## 9. CONCLUSION

We propose an operational approach that scales the throughput of generic matrix multiplication according to the tolerated distortion in the result, expressed as the expected error power instead of the usual worst-case tuning. This can be used in large-scale matrix-based digital signal processing problems that require rapid responses and have certain tolerance to error, or when handling inherently noisy input data. The processor effectively becomes a computation channel controlled in software. Our first results demonstrate that over 170% of the peak performance of a processor can be achieved under stochastic distortion control. For applications that can tolerate imprecision in their computation, this already makes our proposal 45%~180% more efficient than the current state of the art (Figure 6). Further work could investigate the throughput potential of different packing techniques as well as the possibility of incorporating fault detection and correction in the derived computations.

## APPENDIX - PROOFS

*Proof of Proposition 2*: From (4) and under iid input statistics, the resulting output random variable $\hat{\rho}_{m,n}$ is:

$$\hat{\rho}_{m,n} = \sum_{l=0}^{L-1} \hat{\alpha}_{m,l} \hat{\beta}_{l,n} \qquad (28)$$

where, $\forall m, l, n: \mu_{\hat{\alpha}_{m,l}} = \mu_{\hat{\beta}_{l,n}} = 0, \sigma_{\hat{\alpha}_{m,l}} = \sigma_\alpha + \sigma_{\nu_\alpha}, \sigma_{\hat{\beta}_{l,n}} = \sigma_\beta + \sigma_{\nu_\beta}$. We can express $\hat{\rho}_{m,n}$ in affine form [24]:

$$\hat{\rho}_{m,n} = \sum_{l=0}^{L-1} \left( \mu_{\hat{\alpha}_{m,l}} \mu_{\hat{\beta}_{l,n}} + \mu_{\hat{\alpha}_{m,l}} \sigma_{\hat{\beta}_{l,n}} \chi_l + \mu_{\hat{\beta}_{l,n}} \sigma_{\hat{\alpha}_{m,l}} \chi_{L+l} + \sigma_{\hat{\alpha}_{m,l}} \sigma_{\hat{\beta}_{l,n}} \chi_{2L+l} \right) \quad (29)$$

with $\chi_0, \ldots, \chi_{3L-1} \sim P_\chi(1)$ zero-mean iid RVs with unit standard deviation. Expanding on (29), we have:

$$\hat{\rho}_{m,n} = \sum_{l=0}^{L-1} \left( \sigma_\alpha \sigma_\beta \psi_l + \sigma_\alpha \sigma_{\nu_\beta} \psi_{L+l} + \sigma_\beta \sigma_{\nu_\alpha} \psi_{2L+l} + \sigma_{\nu_\alpha} \sigma_{\nu_\beta} \psi_{3L+l} \right) \quad (30)$$

with $\psi_0, \ldots, \psi_{4L-1} \sim P_\psi(1)$ zero-mean iid RVs with unit standard deviation.

The equivalent expression for the inner product of a row of **A** with a column of **B** is:

$$\rho_{m,n} = \sum_{l=0}^{L-1} \sigma_\alpha \sigma_\beta \psi_l . \quad (31)$$

Hence, the noise is expressed by:

$$\rho_{m,n} - \hat{\rho}_{m,n} = \sum_{l=0}^{L-1} \left( \sigma_\alpha \sigma_{\nu_\beta} \psi_{L+l} + \sigma_\beta \sigma_{\nu_\alpha} \psi_{2L+l} + \sigma_{\nu_\alpha} \sigma_{\nu_\beta} \psi_{3L+l} \right) \quad (32)$$

and the expected noise power, $E\{(\rho_{m,n} - \hat{\rho}_{m,n})^2\}$, is given by (19). ∎

*Proof of Proposition 3*: We express $c_\mathbf{B}$ in function of $c_\mathbf{A}$ from (17). We then link $c_\mathbf{A}$ with the expected SNR via (21), given that $E\{(\rho_{m,n} - \hat{\rho}_{m,n})^2\} \equiv L(\sigma_\alpha \sigma_\beta)^2 / 10^{0.1 E\{S_{\text{subblock}}\}}$. Replacing (21) in the last equation and after a few straightforward algebraic manipulations, we reach:

$$\sigma_\alpha^2 c_{\text{tot}}^2 c_\mathbf{A}^4 + \left[ \frac{[12 s(R_{\max}, W)]^2 + 1}{12 c_{\text{tot}}^2} - \frac{12 \sigma_\alpha^2 \sigma_\beta^2}{10^{0.1 E\{S_{\text{subblock}}\}}} \right] c_\mathbf{A}^2 + \sigma_\beta^2 = 0 \quad (33)$$

with $c_{\text{tot}}$ defined by (24). Solving (33) for $c_\mathbf{A}^2$ provides:

$$c_\mathbf{A}^2 = \frac{D_{\text{QvR}} \pm \sqrt{D_{\text{QvR}}^2 - 4 \sigma_\alpha^2 \sigma_\beta^2 c_{\text{tot}}^2}}{2 \sigma_\alpha^2 c_{\text{tot}}^2} \quad (34)$$

with $D_{\text{QvR}}$ representing the quantization-versus-representation noise, defined by (23). From (34), in order for $c_\mathbf{A}$ to be real, $|D_{\text{QvR}}| \geq 2\sigma_\alpha \sigma_\beta c_{\text{tot}}$ and $D_{\text{QvR}} > 0$. This leads to companders defined by (22). ∎

*Proof of Proposition 4*: The term $D_{\text{QvR}}$ of (23) expresses the quantization versus the representation noise. In particular, when quantization is refined via the use of larger $R_{\max}$, the first term of (23), $12\sigma_\alpha^2 \sigma_\beta^2 10^{-0.1 E\{S_{\text{subblock}}\}}$, is monotonically decreasing as $E\{S_{\text{subblock}}\}$ (obtained SNR) is monotonically increasing. However, the second part of (23) is monotonically increasing for increased $R_{\max}$, since $c_{\text{tot}}^{-2}$ is proportional to $R_{\max}^2$ and $s(R_{\max}, W)$ is monotonically increasing with $R_{\max}$ [something that was also verified experimentally in Figure 3(a)]. Hence, the optimal point is found at the value of $R_{\max}$ for which $D_{\text{QvR}} = 2\sigma_\alpha \sigma_\beta c_{\text{tot}}$, i.e. the maximum possible SNR is obtained with companders that remain marginally admissible by (25). This condition leads to the companders shown in (26). Furthermore, under this condition, solving (24) for $E\{S_{\text{subblock}}\}$ derives the form for $E\{S_{\text{subblock}}\}$ given in (27). Since this expression depends on $s(R_{\max}, W)$, for which no analytic model exists, we can solve this equation numerically to derive the optimal (maximum) value for $E\{S_{\text{subblock}}\}$ and $R_{\max}$ under the input data statistics and the given packing $W$. ∎

## REFERENCES

[1] J. Yang, *et al*, "Two-dimensional PCA: A new approach to appearance-based face representation and recognition," *IEEE Trans. on Pattern Anal. and Machine Intel.*, vol. 26, no. 1, pp. 131-137, Jan. 2004.

[2] F. Abdelkefi and P. Duhamel, "Impulsive noise cancellation in multicarrier transmission," *IEEE Trans. On Commun.*, vol. 53, no. 1, pp. 94-106, Jan. 2005.

[3] J. Bilmes, K. Asanovic, C.-W. Chin and J. Demmel, "Using PHiPAC to speed error back-propagation learning," *Proc. IEEE Acoust., Speech, and Signal Process.*, ICASSP, vol. 5, pp. 4153-4156, Apr. 1997.

[4] Q. Du and J. E. Fowler, "Hyperspectral image compression using JPEG2000 and principal component analysis," *IEEE Geosc. and Remote Sens. Lett.*, vol. 4, no. 2, pp. 201-205, Apr. 2007.

[5] R. C. Whaley, A. Petitet, and J. Dongarra, "Automatically empirical optimization of software and the ATLAS project," *Parallel Comp.*, vol. 27, no. 1-2, pp. 3-35, Jan. 2001.

[6] J. L. Gustafson and B. S. Greer, "A hardware accelerator for the Intel Math Kernel," *White Paper*, ClearSpeed Technology Inc., 2006.

[7] T. Yeh *et al*, "Fool me twice: Exploring and exploiting error tolerance in physics-based animation," *ACM Trans. on Computer Graphics*, vol. 29, no. 1, art. 5, Dec. 2009.

[8] J. T. Ludwig, S. H. Nawab, and A. Chandrakasan, "Low-power digital filtering using approximate processing," *IEEE Journal of Solid-State Circ.*, vol. 31, no. 3, pp. 395-340, Mar. 1996.

[9] R. Yuster and U. Zwick, "Fast sparse matrix multiplication," *ACM Trans. on Algorithms*, vol. 1, no. 1, July 2005.

[10] J. Kurzak and J. Dongarra, "Implementation of mixed precision in solving systems of linear equations on the Cell processor," *Concurr. and Comput.: Practice and Exper.*, vol. 19, no. 10, pp. 1371-1385, Jul. 2007.

[11] P. Drineas, *et al*, "Fast monte carlo algorithms for matrices I: Approximating matrix multiplication," *SIAM J. Comput.*, vol. 36, no. 1, pp. 132-157, May 2006.

[12] N. Shanbhag, *et al*, "Stochastic computation," *ACM Design Automation Conf.*, pp. 859-864, June 2010.

[13] D. Anastasia and Y. Andreopoulos, "Software designs of image processing tasks with incremental refinement of computation,"*IEEE Trans. on Image Process*, vol. 19, no. 8, pp. 2099-2114, Aug. 2010.

[14] D. Anastasia and Y. Andreopoulos, "Linear image processing operations with operational tight packing,"*IEEE Signal Process. Letters*, vol. 17, no. 4, pp. 375-378, April 2010.

[15] A. Frank and A. Asuncion, *UCI Machine Learning Repository*, Irvine, CA: University of California, School of Information and Computer Science, *YearPredictionMSD Data Set* (a subset of the data from the Million Song Dataset), http://archive.ics.uci.edu/ml/datasets/YearPredictionMSD .

[16] D. Anastasia and Y. Andreopoulos, "Throughput-precision computation for generic matrix multiplication: toward a computation channel for high-performance digital signal processing," *Proc. IEEE Dig. Sig. Process.*, Corfu, 2011.

[17] *Supplementary material available in the 2nd author's webpage*: http://www.ee.ucl.ac.uk/~iandreop/ORIP.html

[18] D. Lammers, "The era of error-tolerant computing," *IEEE Spectrum*, Nov. 2010.

[19] S. Nassif, *et al*, "A resilience roadmap," *Proc. Design and Test in Europe*, DATE, Dresden, Apr. 2010.

[20] A. Kadyrov and M. Petrou, "The "Invaders" algorithm: range of values modulation for accelerated correlation," *IEEE Trans. Pattern Anal. Machine Intel.*, vol. 28, no. 11, pp. 1882-1886, Nov. 2006.

[21] D. Goldberg, "What every computer scientist should know about floating-point arithmetic," *ACM Computing Surveys*, vol. 23, no. 1, pp. 5-47, March 1991.

[22] K. Goto and R. A. Van de Geijn, "Anatomy of high-performance matrix multiplication," *ACM Trans. on Math. Software*, vol. 34, no. 3, Article 3, May 2008.

[23] Y.-C. Lin and S.-C. Tai, "Fast full-search block-matching algorithm for motion-compensated video compression," *IEEE Trans. On Commun.*, vol. 45, no. 5, pp. 527-531, May 1997.

[24] L. H. de Figueiredo and J. Stolfi, "Affine arithmetic: concepts and applications," *Numer. Algorithms*, vol. 37, pp. 147-158, 2004.